\documentclass[manuscript]{aastex}
\usepackage{natbib}

\shorttitle{Electron acceleration at super-high Mach shock}
\shortauthors{Matsumoto and Hoshino}

\begin{document}


\title{ELECTRON ACCELERATIONS AT HIGH MACH NUMBER SHOCKS: TWO-DIMENSIONAL PARTICLE-IN-CELL SIMULATIONS IN VARIOUS PARAMETER REGIMES}


\author{Yosuke Matsumoto}
\affil{Department of Physics, Chiba University, Yayoi-cho 1-33, Inage-ku, Chiba 263-8522, Japan}
\and
\author{Takanobu Amano and Masahiro Hoshino}
\affil{Department of Earth and Planetary Science, University of Tokyo, Hongo 1-33, Bunkyo-ku, Tokyo 113-0033, Japan}
\email{ymatumot@astro.s.chiba-u.ac.jp}




\begin{abstract}
Electron accelerations at high Mach number collision-less shocks are investigated by means of two-dimensional electromagnetic Particle-in-Cell simulations with various Alfv\'en Mach numbers, ion-to-electron mass ratios, and the upstream electron $\beta_e$ (the ratio of the thermal pressure to the magnetic pressure). We found electrons are effectively accelerated at a super-high Mach number shock ($M_A\sim30$) with a mass ratio of $M/m=100$ and $\beta_e=0.5$. The electron shock surfing acceleration is an effective mechanism for accelerating the particles toward the relativistic regime even in two dimensions with the large mass ratio. Buneman instability excited at the leading edge of the foot in the super-high Mach number shock results in a coherent electrostatic potential structure. While multi-dimensionality allows the electrons to escape from the trapping region, they can interact with the strong electrostatic field several times. Simulation runs in various parameter regimes indicate that the electron shock surfing acceleration is an effective mechanism for producing relativistic particles in extremely-high Mach number shocks in supernova remnants, provided that the upstream electron temperature is reasonably low.

\end{abstract}



\section{Introduction}
Origin of cosmic rays has been a longstanding issue in astrophysics. A rapid progress in understanding their accelerators has been made by observations of radio and X-ray synchrotron emissions from the shell of supernova remnants (SNRs)  \citep{Koyama_1995,Winkler_Long_1997,Bamba_2003,Long_2003}. Those observations have indicated that electrons are strongly accelerated up to TeV energies out of thermal population in the vicinity of non-relativistic collision-less shocks.

Diffusive shock acceleration (DSA) is a widely accepted theory responsible for particle acceleration at shocks, which nicely predicts a power low energy spectrum \citep[e.g.,][]{Blandford_Eichler_1987}. This theory has achieved many successes in explaining energized ions, for example, at the earth's bow shock \citep{Scholer_1980}. However, it cannot be straightforwardly applied to electron accelerations at SNRs: In DSA process, particles are scattered by MHD waves back and forth across the shock. The process is thus unlikely to occur for thermal electrons whose gyro radius is much smaller than the wavelength of the MHD waves or the shock width (ion inertia length). They are strongly tied to the magnetic field line and convect toward downstream adiabatically. This small-scale property of the electrons makes it challenging to explain the electron accelerations by DSA at SNR shocks. A pre-acceleration mechanism for the electrons that enables DSA to work subsequently has been proposed for the last decade in order to overcome this difficulty, which is now known as the injection problem.

In supercritical quasi-perpendicular shocks ($M_A > 2.7$, where $M_A$ is the Alfv\'en Mach number), it has been pointed out that plasma microscopic instabilities play important roles in energizing electrons  \citep{Papadopoulos_1988,Cargill_Papadopoulos_1988,Scholer_2003,Scholer_Matsukiyo_2004}. In the particular case of high Mach number shocks, electrostatic instabilities arise due to velocity difference between the reflected ions and the incident (upstream) ions and electrons \citep{Papadopoulos_1988,Cargill_Papadopoulos_1988}. Buneman instability is initially excited by the reflected ion beam interacting with the upstream electrons. The instability strongly heats the electrons perpendicular to the magnetic field, which then gives a favorable condition for destabilizing the ion acoustic instability in the foot region. Particle-in-Cell (PIC) simulations, which follow ion and electron motions along with the electromagnetic fields development in a self-consistent manner, have confirmed those ideas. Furthermore, a rapid acceleration of the electrons by the convective electric field accompanied with particle trapping by the Buneman instability was found to contribute to nonthermal electron production \citep{Shimada_Hoshino_2000,MacClements_2001,Hoshino_Shimada_2002,Shimada_2010}. This electron shock surfing acceleration has been considered to be a very efficient mechanism at extremely-high Mach number shocks such as SNR shocks that injects thermal electrons to DSA.

It is only recently that evolutions of non-relativistic shocks has been investigated by means of two-dimensional PIC simulations \citep{Umeda_2009,Amano_Hoshino_2009a,Lembege_2009,Kato_Takabe_2010,Riquelme_Spitkovsky_2011}. The efficiency of the electron shock surfing acceleration in multi dimensions has been discussed in different physical parameter regimes, such as the ion-to-electron mass ratio, Alfv\'en Mach number, upstream electron $\beta_e$ (the ratio of the thermal pressure to the magnetic pressure), and magnetic field configurations; the two-dimensional PIC simulation still requires large computational resources and only allows limited simulation runs. Consequently, the results lead to different conclusions. The simulation study with a relatively low mass ratio ($M/m=25$) concluded that the shock surfing acceleration is an effective mechanism even in two dimensions in a high ($M_A=14$) Alfv\'en Mach number shock \citep{Amano_Hoshino_2009a}. \citet{Kato_Takabe_2010} examined an extremely-high Alfv\'en Mach number ($M_A=130$) shock with a high upstream electron $\beta_e$ condition. In their simulation result, the downstream electron energy spectrum exhibited almost a Maxwellian. They attributed the lack of the nonthermal population to the in-plane magnetic field configuration which was different from the out-of-plane configuration adopted by \citet{Amano_Hoshino_2009a}. \citet{Riquelme_Spitkovsky_2011} conducted simulation runs with various mass ratios and concluded that the Buneman instability is suppressed in a large mass ratio case ($M/m=400$) under the condition of $M_A=7$ and the electron shock surfing acceleration will not be realized at SNR shocks. Those contrary conclusions regarding the electron shock surfing acceleration at SNRs are mainly due to the limited choices of the mass ratio, Alfv\'en Mach number, and the upstream electron $\beta_e$, which should be carefully chosen based on the theoretical estimations \citep{Papadopoulos_1988,Matsukiyo_2010}.

In this paper, we present two-dimensional electromagnetic PIC simulations of collision-less shocks with various mass ratios, Alfv\'en Mach numbers, and the upstream electron $\beta_e$ conditions in order to clarify the efficiency of the electron shock surfing acceleration at SNR shocks. The parameters are carefully chosen in accordance with the linear and nonlinear theories of the Buneman instability in the foot region, which are reviewed in Section 2. In Section 3, the numerical setup as well as our two-dimensional PIC code are described. In Section 4, we first overview a simulation result in which the maximum energy gain of the electrons is observed. Electron acceleration processes are analyzed in detail. Then dependence of the acceleration mechanisms on the physical parameters are discussed. The results are summarized and discussed in the last section.

\section{Unstable condition and the saturation level of Buneman instability in the foot region}
The electron shock surfing acceleration is basically supported by the electrostatic field excited by the Buneman instability at the leading edge of the foot region. The free energy of the instability is the relative drift motion between the incoming electrons and the reflected ions. The region becomes unstable when the relative speed exceeds the thermal speed of the electron, $\Delta V/v_{the}>1$. 
In applying the unstable condition to the foot region, the velocity of the incoming ion and electrons, and the reflected ions are expressed as
\begin{eqnarray}
V_i = V_0,\\ 
V_r = -V_0,\\
V_e = \frac{1-\alpha}{1+\alpha}V_0,
\end{eqnarray}
by assuming the zero net current condition
\begin{equation}
n_0 V_i+n_r V_r-(n_0+n_r) V_e = 0,
\end{equation}
where $n_0$ is the upstream number density, $n_r$ is the number density of the reflected ion, $V_0$ is the upstream speed, $V_i$ is incoming ion speed, $V_e$ is the incoming electron speed, and $V_r$ is the reflected ion speed. $\alpha=n_r/n_0$ is the density ratio of the reflected and incoming ions, which is typically $\sim 0.2$ for high Mach number shocks \citep{Papadopoulos_1988,Leroy_1981,Leroy_1982,Leroy_1983}. Using $\Delta V = V_e - V_r$, the unstable condition reads as
\begin{equation}
M_A \ge \frac{1+\alpha}{2}\sqrt{\beta_e} \left(\frac{M}{m}\right)^{\frac{1}{2}},
\label{eqn:unstable_condition}
\end{equation}
where $M_A$ is the Alfv\'en Mach number, $M$ and $m$ are the ion and electron rest mass, and $\beta_e$ is the ratio of the electron pressure to the magnetic pressure (plasma beta).

Given the Buneman instability being excited, the electrons are accelerated by the convective electric field while they are trapped in the electrostatic potential well. The maximum energy gain is determined by force balance between the trapping force by the saturated Buneman instability ($qE_{BI}$) and the Lorentz force ($qvB_0/c)$ for escaping. To be accelerated toward relativistic speeds ($v \sim c$), the electric force must be larger than the Lorentz force \citep{Hoshino_Shimada_2002}:
\begin{equation}
\frac{E_{BI}}{B_0} \ge 1.
\label{eqn:eb_ratio}
\end{equation}
The saturation level of the Buneman instability ($|E_{BI}|$) can be estimated by introducing an energy conversion rate $C$ of the electron drift energy density as
\begin{equation}
\frac{E_{BI}^2}{8\pi} = \frac{1}{2}m n_0 \Delta V^2 C. \\
\label{eqn:conversion}
\end{equation}
The conversion rate $C$ has been analyzed by the nonlinear theory of the Buneman instability \citep{Ishihara_1980} and found to depend weakly on the mass ratio $C\sim (m/M)^{1/3}$. In multi dimensions, the saturation level is reduced to $\sim 25\%$ of above estimate due to the resonant wave-particle interactions with waves oblique to the beam direction \citep{Amano_Hoshino_2009b}. Using $\Delta V = V_e - V_r$ and eq.(\ref{eqn:conversion}), and taking account of the multidimensionality, eq. (\ref{eqn:eb_ratio}) leads to
\begin{equation}
M_A \ge (1+\alpha)\left(\frac{M}{m}\right)^{\frac{2}{3}}.
\label{eqn:saturation_level}
\end{equation}

Eq. (\ref{eqn:unstable_condition}) indicates that the electron shock surfing acceleration occurs at a sufficiently high Alfv\'en Mach number shock in a low beta plasma. Furthermore, for non-thermal relativistic electrons to be observed, eq. (\ref{eqn:saturation_level}) shows that the mass ratio restricts Alfv\'en Mach number more severely with a power of $2/3$ rather than $1/2$ in the unstable condition, indicating a need for numerical simulations of super-high Alfv\'en Mach number shocks with larger mass ratios, for instance, $M_A>26$ for $M/m=100$. In other words, simulations with artificial small mass ratios would easily give an unrealistic electron acceleration in relatively low Alfv\'en Mach number shocks.

\section{Numerical setup}
Two-dimensional electromagnetic Particle-in-Cell simulation code is used for examining shock evolutions. The code is parallelized via domain decomposition by using Message Passing Interface (MPI) library and OpenMP and runs efficiently on massively parallel supercomputer systems.

Particles are continuously injected from one boundary ($x=0$) with a speed $V_0$ toward the other end ($x=L_x$) where the particles are reflected (injection method), resulting in the shock formation that propagates toward $x=0$. Number density in the upstream is 40 particles per cell for each species (ion and electron). The injected plasma carries the z component of the magnetic field ($B_0$) and the convective electric field $E_y = E_0=V_0 B_0$. Thus we deal with purely perpendicular shocks in the downstream-rest frame. The periodic boundary condition is applied in the y direction. Simulation box sizes in the x and y directions are $L_x=12\ V_0/\Omega_{gi} = 12\ M_A \lambda_i$ and $L_y=5\ \lambda_i$, respectively, where $M_A = V_0/V_A$, $\Omega_{gi}=qB_0/Mc$ is the ion gyro frequency, $\lambda_i=c/\omega_{pi}=c\sqrt{m/4\pi n_0 q^2}$ is the ion inertia length, $V_A=B_0/\sqrt{4 \pi M n_0}$ is the Alfv\'en speed, and $n_0$ is the number density in the upstream region (40 particles per cell). Other notations of $c$ and $q$ respectively represent the speed of light and elementary charge. The grid size $\Delta h$ and the time step size $\Delta t$ are set as $\Delta h = \lambda_D$ and $\omega_{pe} \Delta t=0.025$, where $\lambda_D = \sqrt{T_e/4 \pi n_0 q^2}$ is Debye length in the upstream region, $T_e$ is the electron temperature, and $\omega_{pe}=\sqrt{4 \pi n_0 q^2/m}$ is the electron plasma frequency. In the following, we use the units of $\Omega_{gi}^{-1}$, $c/\omega_{pi}$, $n_0$, $B_0$, $c$, and $mc^2$ for time, space, number density, electromagnetic fields, velocity, and energy, respectively, unless otherwise stated.

We have carried out several runs with various ion to electron mass ratios ($M/m$), the Alfv\'en Mach numbers ($\hat{M_A}=V_0/V_A$), and the electron plasma $\beta_e=8\pi n_0T_e/B_0^2$, while the ratio of the electron plasma to gyro frequencies ($\omega_{pe}/\Omega_{ge}$) and the temperature ratio of the ion to the electron ($T_i/T_e$) are fixed to $\omega_{pe}/\Omega_{ge}=10.0$ and $T_i/T_e=1$, respectively. Two simulation runs use the mass ratios of 25 and 100 for fixed values of $\hat{M_A}=10$ (respectively $M_A \sim 14$ and $16$ in the shock-rest frame) and $\beta_e=0.5$ in order to examine the mass ratio dependence of the electron acceleration (Runs A and B). Parameters in Run A are the same as used those by \citet{Amano_Hoshino_2009a}. Another simulation run uses $M/m=100$ with $\hat{M_A}=20\ (M_A \sim 30)$ for comparison with the lower Alfv\'en number cases (Run C). We also examined a hot electron case with $\beta_e=4.5$ for comparison with the high Alfv\'en Mach number and a lower electron temperature case (Run D). We totally examined four simulation runs (Runs A-D) with various $M_A$, $M/m$, and $\beta_e$ to assess importance of these parameters to electron acceleration mechanisms. The upstream parameters of the four simulation runs are summarized in Table \ref{tbl:param}. The largest computational resources are used in Run C, in which $5\times10^{9}$ particles are followed in a simulation domain with $24001 \times 1024$ grid points.

\section{Simulation Results}
\subsection{Electron acceleration in Run C}
\subsubsection{Overview of 2D shock structure}
We first present an overview of a two-dimensional perpendicular shock structure in Run C in which the maximum energy gain of the electron was observed. In the following results, the data at a time of $\Omega_{gi} T=6$ is used, after which the energy spectrum of the electron in the downstream region is not significantly changed.

\begin{figure}
\epsscale{0.8}
\plotone{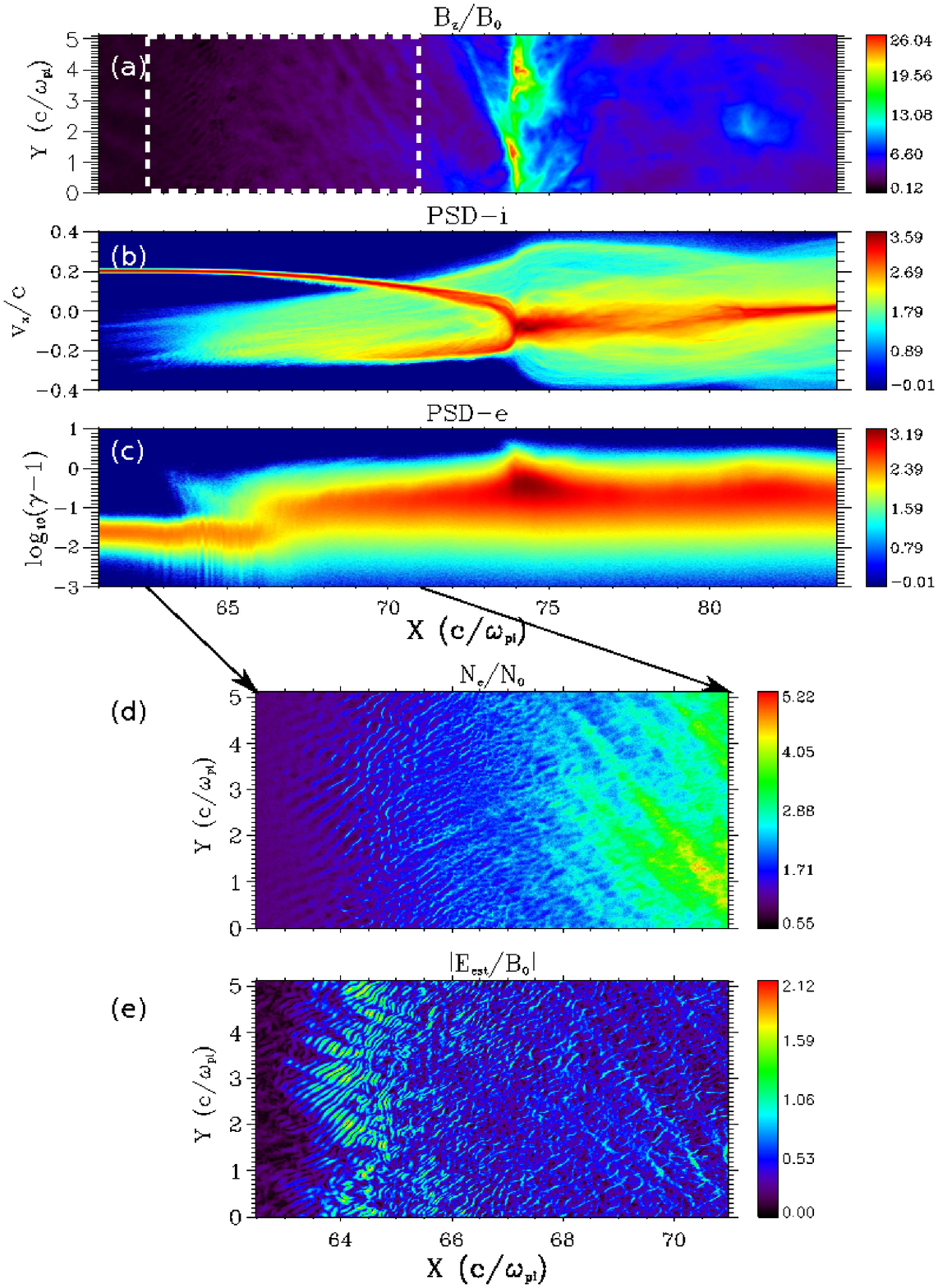}
\caption{Overview of a 2D shock structure in Run C obtained at $\Omega_{gi}T=6$. (a) Two-dimensional profile of $B_z$. (b) Ion phase space (X-vx) plot. The color code shows number of particles in a logarithmic scale. (c) Electron phase space (X-energy) plot. The energy ($\gamma-1$) in the y axis is in a logarithmic scale. (d) Electron number density profile and (e) strength of the electrostatic field (rotation-free-part of the electric field) in the zoomed-up region surrounded by the white-dashed lines in panel (a).}
\label{fig:ov}
\end{figure}

Figure \ref{fig:ov}(a) shows a two-dimensional spatial profile of the z component of the magnetic field. The structure is characterized by the foot region which extends from $X=63\ c/\omega_{pi}$ to the ramp and the overshoot around $X=74\ c/\omega_{pi}$, and the downstream region in $X>74\ c/\omega_{pi}$. The width of the foot (the shock transition region) is well scaled by the gyro radius of the reflected ion ($L \sim 0.7\ V_0/\Omega_{gi}$) \citep{Leroy_1983,Shimada_2010}.

Figure \ref{fig:ov}(b) shows the phase space plot for ions. The ions in the foot consist of the incoming cold population and the relatively hot ion moving toward the upstream region which is specularly reflected by the shock potential at $X\sim74\ c/\omega_{pi}$. At the leading edge of the reflected ion, electrons are preferentially accelerated immediately after entering the foot as seen in the energy-space diagram in Figure \ref{fig:ov}(c). Just behind the region around at $X\sim 63\ c/\omega_{pi}$, the maximum energy of the electrons increased to $\gamma \sim 2$. Then they are further heated in $X > 66.5\ c/\omega_{pi}$ in the foot. In the vicinity of the overshoot, the electron energy again increased up to $\gamma \sim 9$.

Zoomed-up spatial profiles surrounded by the white dashed line in Figure \ref{fig:ov}(a) are shown in Figures \ref{fig:ov}(d) and (e). At the edge of the foot, we can identify fine scale structures of the electron number density in Figure \ref{fig:ov}(d), whose spatial size is much smaller than the ion inertia scale, i.e., the electron scale that is embedded in the MHD scale shock structure. This fine structure is accompanied by electrostatic fields. Figure \ref{fig:ov}(e) shows the electrostatic field strength $|E_{est}| = |-{\bf \nabla}\phi|$ given by solving the Poisson equation of $\nabla^2 \phi = -{\bf\nabla} \cdot {\bf E}$. The amplitude reaches twice as large as the background magnetic field ($B_0$), which is much larger than the upstream convective electric field $|E_y/B_0|=V_0/c=0.2$. Coherent wave trains lie in the x-y plane rather than exhibiting two-dimensional isolated structures \citep{Amano_Hoshino_2009a}. This strong electrostatic field at the edge of the foot is caused by an linear instability between the incoming electron and the reflected ion, which one usually refers to as Buneman instability. The Buneman instability heats the electrons perpendicular to the magnetic field and sets a favorable condition for the ion-acoustic instability, which further energized the electrons in $X > 66.5\ c/\omega_{pi}$ in the foot.

\begin{figure}
\plotone{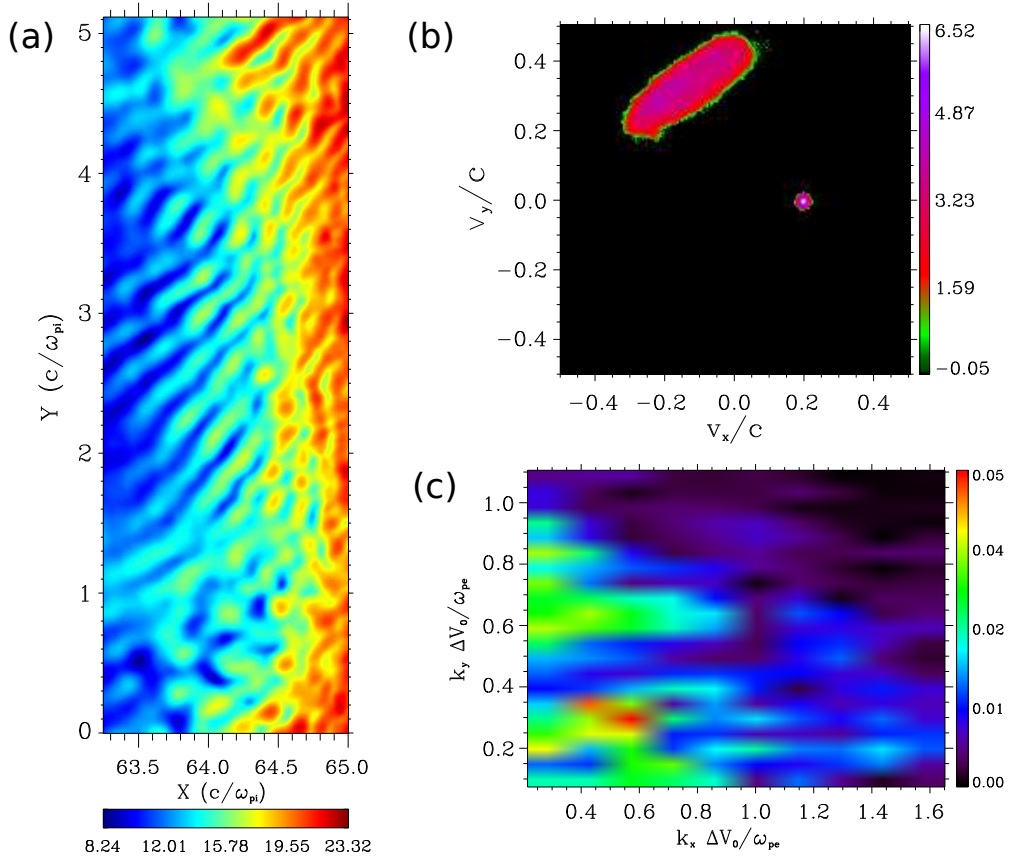}
\caption{(a) Zoomed-up spatial profile of the electrostatic potential energy. The color code is normalized by the upstream  kinetic energy. (b) Velocity distribution function of the ions sampled in the region in panel (a). (c) Fourier power spectrum of the electrostatic potential in two dimensions. Wave number in each axis ($k_x$ and $k_y$) is normalized by $\Delta V_0/\omega_{pe}$, where $\Delta V_0=0.4c$}
\label{fig:bi}
\end{figure}

In order to identify the instability in the two-dimensional plane as the Buneman instability, we focus on the closed up region of interest in Figures \ref{fig:ov}(d) and (e). Figure \ref{fig:bi}(a) shows the zoomed up profile of the electrostatic potential energy $q\phi$. In this region, the cold upstream and the reflected gyrating ions constitute the distribution function shown in Figure \ref{fig:bi}(b). The number of reflected ions with respect to the upstream population is $n_r/n_0=\alpha=0.22$ as expected in previous high Mach number shock simulations \citep{Leroy_1982,Leroy_1983}.

The relative drift speed between the two populations is $\Delta V \sim 0.4c$. This velocity difference is much larger than the upstream electron thermal speed of $0.07c$, thus satisfying the unstable condition of the Buneman instability (eq. (\ref{eqn:unstable_condition})). Figure \ref{fig:bi}(c) shows the Fourier amplitude of the electrostatic potential of Figure \ref{fig:bi}(a) in the $k_x$-$k_y$ space in which Hann window is applied in the x direction. One can identify a strong peak appearing around $(k_x,k_y) = (0.6,0.3)\ \omega_{pe}/\Delta V$. This peak corresponds to the coherent wavy structure that obliquely lies in the x-y plane and $|k|=\sqrt{k_x^2+k_y^2}=0.7\ \omega_{pe}/\Delta V$ agrees with the fastest growing mode at $k_x \Delta V/\Omega_{pe} \sim 1$ predicted by the linear analysis of the Buneman instability \citep{Amano_Hoshino_2009b}.

\subsubsection{Electron acceleration mechanisms}
\begin{figure}
\plotone{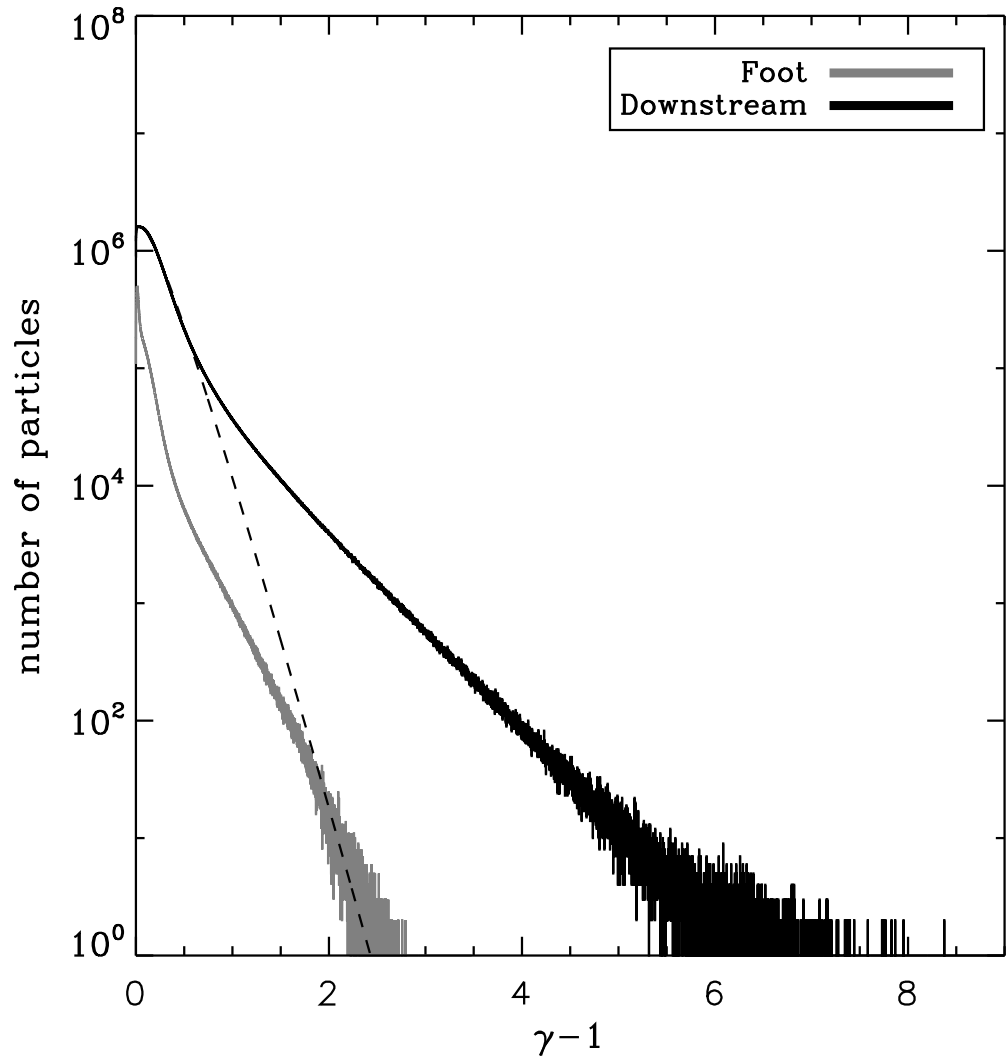}
\caption{Energy spectra of the electron in the foot (gray) and downstream (black) regions obtained in Run C. Dashed line is a fitted Maxwell distribution to the downstream spectrum.}
\label{fig:edist_runc}
\end{figure}

In Run C, a high energy tail of the electron energy distribution function which extends to $\gamma\sim9$ is observed. Figure \ref{fig:edist_runc} shows the energy spectra of the electron in the foot and downstream regions. Particles are sampled in regions $64\ c/\omega_{pi} \le X \le 71\ c/\omega_{pi}$ for the foot spectrum and $74\ c/\omega_{pi} \le X \le 88\ c/\omega_{pi}$ for the downstream spectrum. In the foot region, non-thermal electrons are already produced. The energy distribution consists of the cold, non-accelerated and accelerated populations. The highest energy in this region is $\gamma \sim 3.5$. In the downstream region, a clear two-component distribution is found. One is the thermal electrons whose temperature is $T_e/mc^2=0.13$ and the other consists of accelerated particles whose energy reach $\gamma \sim 9$. Acceleration mechanisms responsible for these non-thermal electrons are found in two ways which are described in the following.

\begin{figure}
\epsscale{0.75}
\plotone{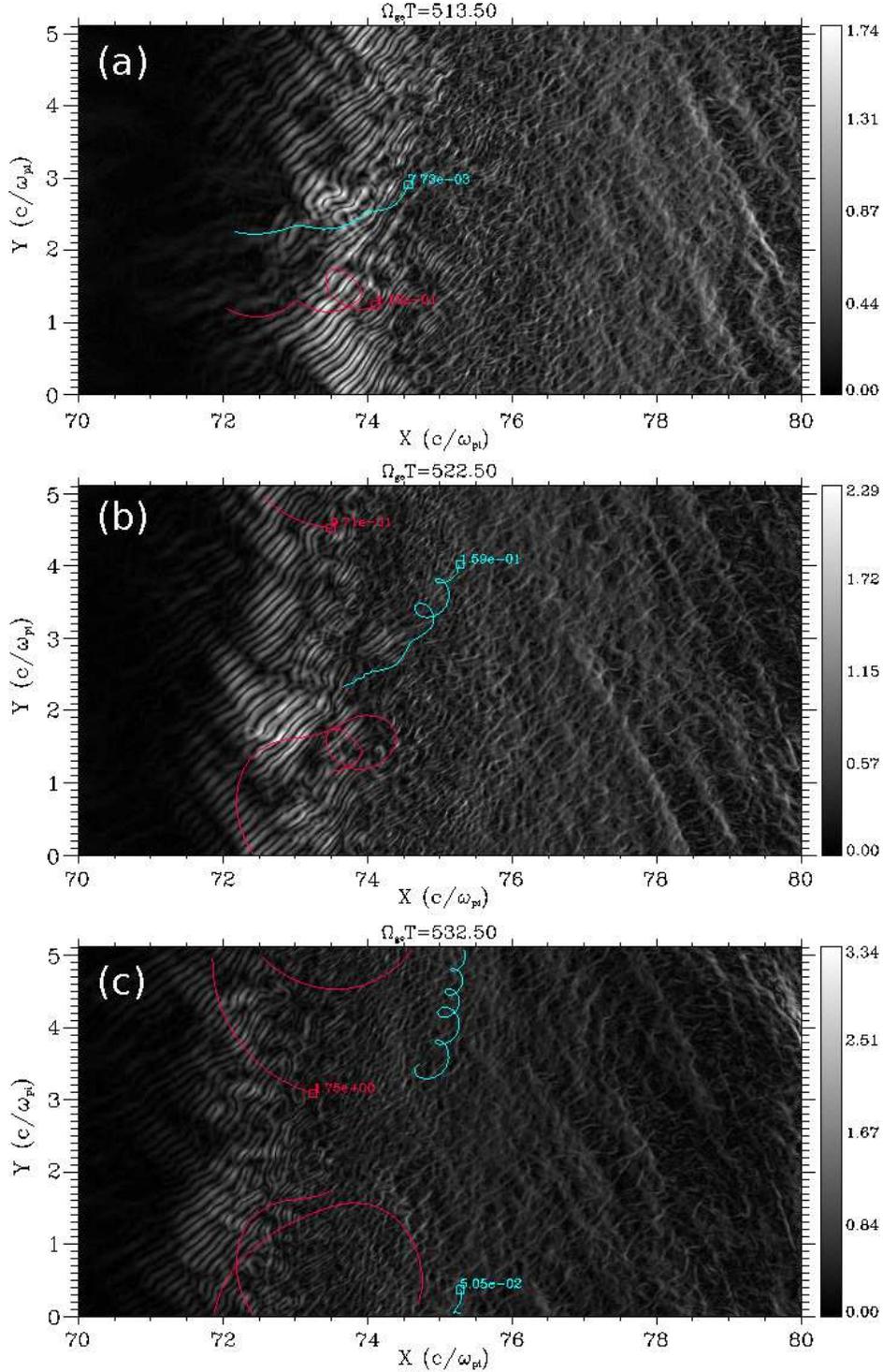}
\caption{Trajectories of accelerated (red) and non-accelerated (cyan) electrons superposed on the two-dimensional profile of the electrostatic field strength. Open square locates the position of the particle at each time step following the trajectory back for $15\ \Omega_{ge}^{-1}$. A number located beside the open square denotes $\gamma-1$. Snapshots are taken at times of (a) $\Omega_{ge}T=513.5$, (b) $\Omega_{gi}T=522.5$, and (c) $\Omega_{gi}T=532.5$, respectively. This figure is also available as an mpeg animation in the electronic edition of the {\it Astrophysical Journal}.}
\label{fig:orbe1}
\end{figure}

First, the electrons are accelerated at the edge of the foot by the shock surfing acceleration as manifested in Figures \ref{fig:ov} and \ref{fig:bi}. When a particle entered the unstable region of the Buneman instability, it is accelerated by the convective electric field while being captured by the electrostatic potential. The present two-dimensional simulation shows in Figure \ref{fig:orbe1}(a) that a sampled electron can be escaped from the potential well at a time after gaining a fraction of energy and drifts toward the downstream. Since the length of the potential well is limited to the order of the ion inertia length, the particle can escape from the well even though the electric field strength is sufficiently large. This is the multi-dimensionality effect. Nevertheless it can enter the unstable region from the downstream-side again since the gyro radius of the accelerated particle becomes larger than before. Then the second surfing acceleration is realized. In the trajectory in Figure \ref{fig:orbe1}(b), the energy gain is more effective than in the previous encounter, and the energy reached $\gamma \sim 2$. The third encounter with the unstable region is possible in the same manner, and the energy is increased to $\gamma \sim 3$ (Figure \ref{fig:orbe1}(c)). For accelerated particles, they typically experience the shock surfing acceleration three times as shown in this example. The present acceleration mechanism can be applied to lucky particles that will constitute the non-thermal population, and most of the particles that will constitute the thermal population drift toward the downstream without being captured as shown with cyan lines in Figure \ref{fig:orbe1}. Nevertheless the shock surfing acceleration is so effective as it can energize the electrons from $\gamma \sim 1$ to the relativistic regime of $\gamma \sim 3$ within a time scale of $\sim20\ \Omega^{-1}_{ge}$, or $\sim 0.2\ \Omega^{-1}_{gi}$ in this particular case with $M/m=100$.

\begin{figure}
\plotone{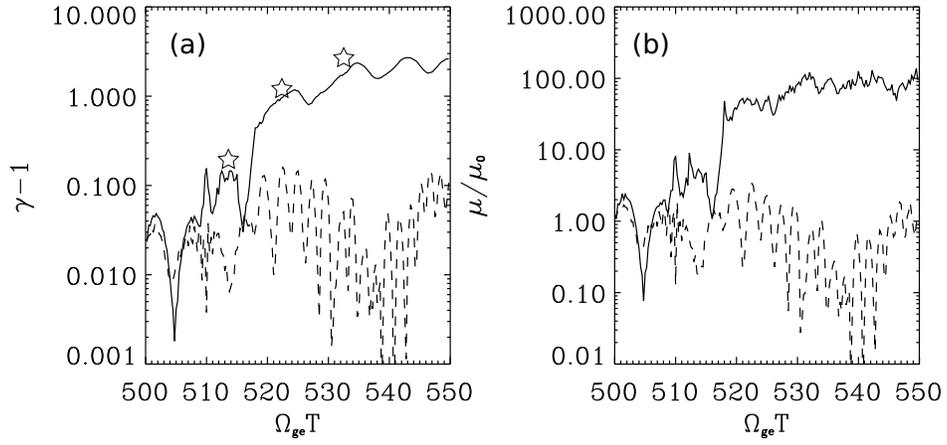}
\caption{Time histories of (a) the energy ($\gamma-1$) and (b) the magnetic moment. Solid and dashed lines are for the accelerated and non-accelerated particles, respectively. Stars in panel (a) indicate the times in Figure \ref{fig:orbe1}.}
\label{fig:ehist1}
\end{figure}

Time histories of the energy and the magnetic moment (the first adiabatic invariant) normalized by $U_0^2/2B_0=V_0^2/2(1-V_0^2/c^2)B_0$ are shown in Figure \ref{fig:ehist1}. The energy history of the accelerated particle is threefold corresponding to Figures \ref{fig:orbe1}(a)-(c). The maximum energy gain is found at $\Omega_{ge}T\sim517$ when $\gamma-1$ is rapidly increased from $0.05$ to $1.0$ within a time scale of $5\ \Omega^{-1}_{ge}$ (Figure \ref{fig:ehist1}(a)). Since the shock surfing acceleration is due to the direct acceleration by the DC electric field, the magnetic moment also rapidly increases as the energy is increased (Figure \ref{fig:ehist1}(b)). After the third surfing acceleration at $\Omega_{ge}T\sim530$, the energy and the magnetic moment are kept constant until $\Omega_{ge}T=550$. On the other hand, the non-accelerated particle remains at the initial energy level while the normalized magnetic moment fluctuates around one; neither obvious non-adiabatic nor adiabatic accelerations works on this particle as it travels in the foot region.

\begin{figure}
\plotone{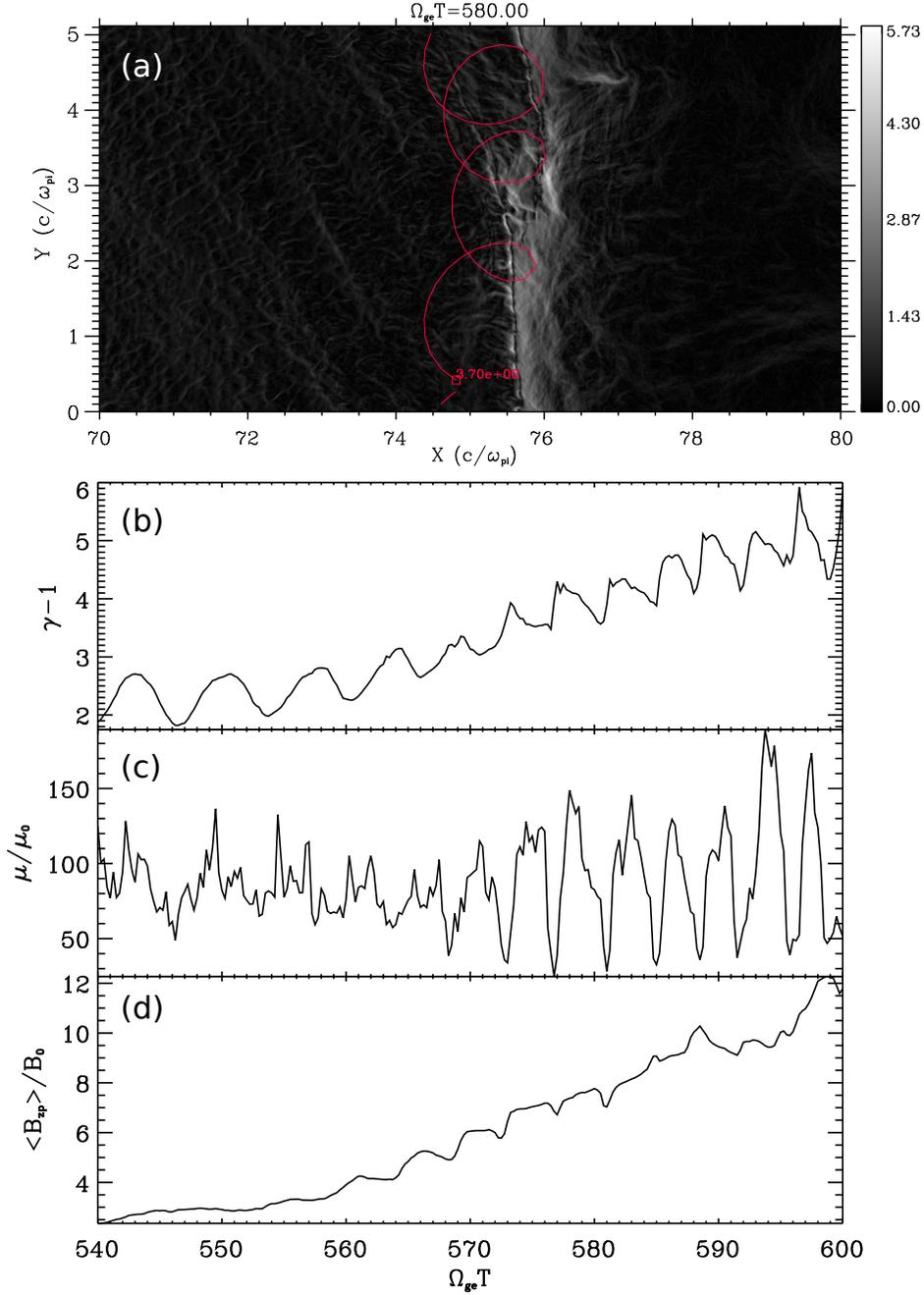}
\caption{(a) Trajectory of the accelerated particle in the same format as Figure \ref{fig:orbe1}. Snapshot is taken at $\Omega_{ge}T = 580$. This figure is also available as an mpeg animation in the electronic edition of the {\it Astrophysical Journal}. Time histories of (b) $\gamma-1$, (c) the magnetic moment, and (d) $B_z$ at the particle's position. Running average for $15\ \Omega_{ge}^{-1}$ is applied to the magnetic field profile (d).}
\label{fig:orbe2}
\end{figure}

The maximum energy achieved by the surfing acceleration at the edge of the foot is up to $\gamma\sim3$, which is not so sufficiently large as the maximum energy of $\gamma \sim 9$ in the downstream (see also Figure \ref{fig:edist}). An additional acceleration process is also found as the particle approached the shock surface. Figure \ref{fig:orbe2}(a) shows a trajectory of the accelerated electron in Figure \ref{fig:orbe1} at $\Omega_{ge}T=580$ after the first acceleration process. The particle undergoes the grad-B drift in the -y direction while being deflected at the overshoot. Figure \ref{fig:orbe2}(b) shows that the acceleration starts when the particle encountered with the shock surface at $\Omega_{ge}T\sim560$. After this time, the energy variation exhibits a sawtooth pattern rather than a sinusoidal one in $\Omega_{ge}T=540-560$; the energy increases rapidly and decays gradually. During the acceleration, the magnetic moment stays at a certain level on average, but is highly oscillatory (Figure \ref{fig:orbe2}(c)). Figure \ref{fig:orbe2}(d) shows the running average of the magnetic field at the particle's position. The particle experiences larger magnetic fields as it is energized in time. These overall signatures in Figures \ref{fig:orbe2}(b)-(d) indicate that the acceleration at the shock surface is basically an adiabatic process.

\begin{figure}
\plotone{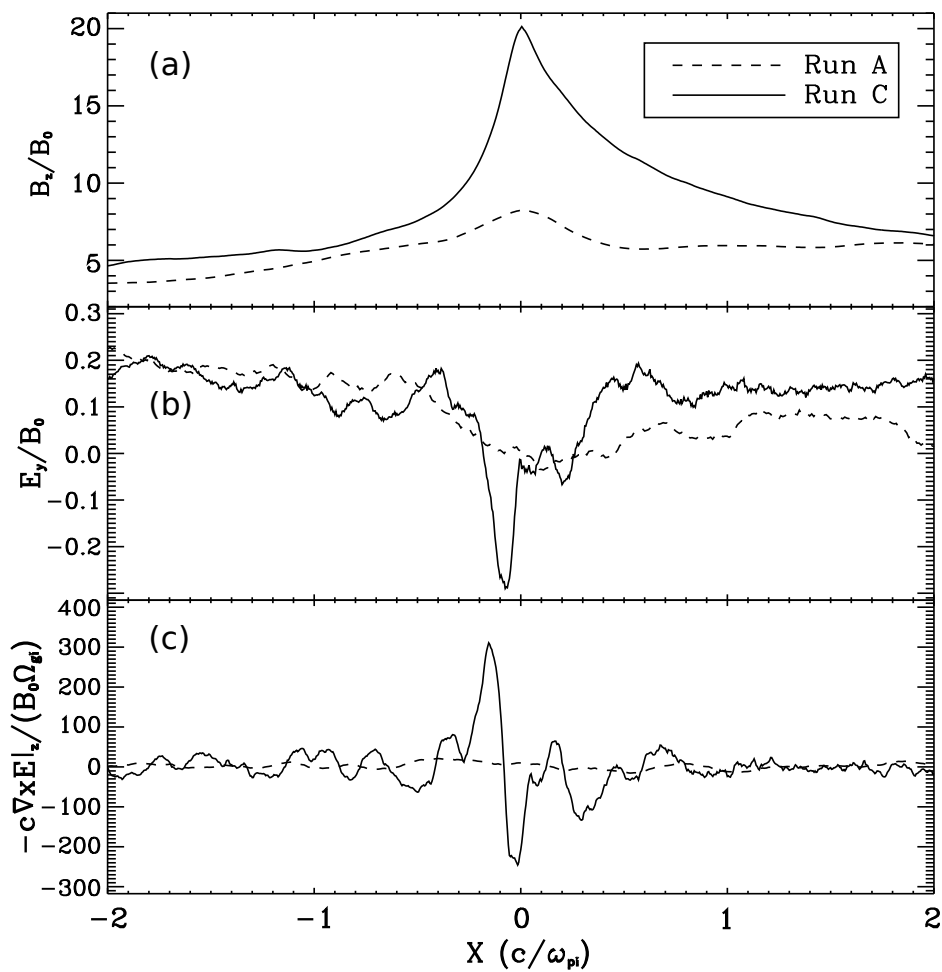}
\caption{Comparisons of (a) $B_z$, (b) $E_y$, and (c) $-{\bf c\nabla}\times {\bf E}|_z$ around the overshoot between Run A (dashed line) and Run C (solid line). The spatial profile across the shock is averaged in the y direction with respect to the overshoot position at each y position.}
\label{fig:eb_compar}
\end{figure}

Figures \ref{fig:eb_compar}(a) and \ref{fig:eb_compar}(b) shows the magnetic and electric fields' profiles with respect to the overshoot position averaged in the y direction (solid lines). The region in the vicinity of the overshoot is enlarged. The electric field is calculated in the shock-rest frame based on the electron frozen-in condition (the generalized Ohm's law including the hall term) as
\begin{equation}
E_y^{'}(x) = E_y(x) -V_{sh}B_z(x),
\end{equation}
where $V_{sh}$ is the electron convective velocity in the x direction at the overshoot position. The amplitude of the magnetic field at the overshoot reached twenty times as large as the upstream value within a scale of the ion inertia length, giving rise to the fast grad-B drift speed of the relativistic electron. Since the direction of the convective electric field is opposite to the drift motion, the electron can gain energy as it travels in the y direction. This is similar to the typical (ion) shock drift acceleration, but for the relativistic electron \citep{Armstrong_1985,Krauss-Varban_1989}. Furthermore, the electron is accelerated by the strong negative $E_y$ in the ramp ($X = -0.1\ c/\omega_{pi}$) unlike the typical shock drift acceleration. This electric field is induced by time variations of the shock structure, that is, the shock reformation \citep{Lembege_Savoini_1992,Scholer_2003,Lembege_2009}. Figure \ref{fig:eb_compar}(c) shows the z component of $-{\bf \nabla} \times {\bf E}=\partial {\bf B} / \partial t$. It takes large positive and negative values in the ramp in the shock-rest frame, indicating that a new shock front is forming ahead of the existing overshoot position. Since the inductive electric field strength is larger than the convective electric field ($|E_{ind}| \sim 0.3B_0$), the betatron acceleration is also effective for the relativistic electron in the present simulation run.

\subsection{Mass ratio and Alfv\'en Mach number dependencies}
We have shown the shock surfing acceleration and additional adiabatic accelerations at the shock surface in Run C in which the maximum energy gain of the electron is observed. Here we examine these mechanisms in the other simulation runs with various ion-to-electron mass ratios ($M/m$) and Alfv\'en Mach numbers ($M_A$).

\begin{figure}
\plotone{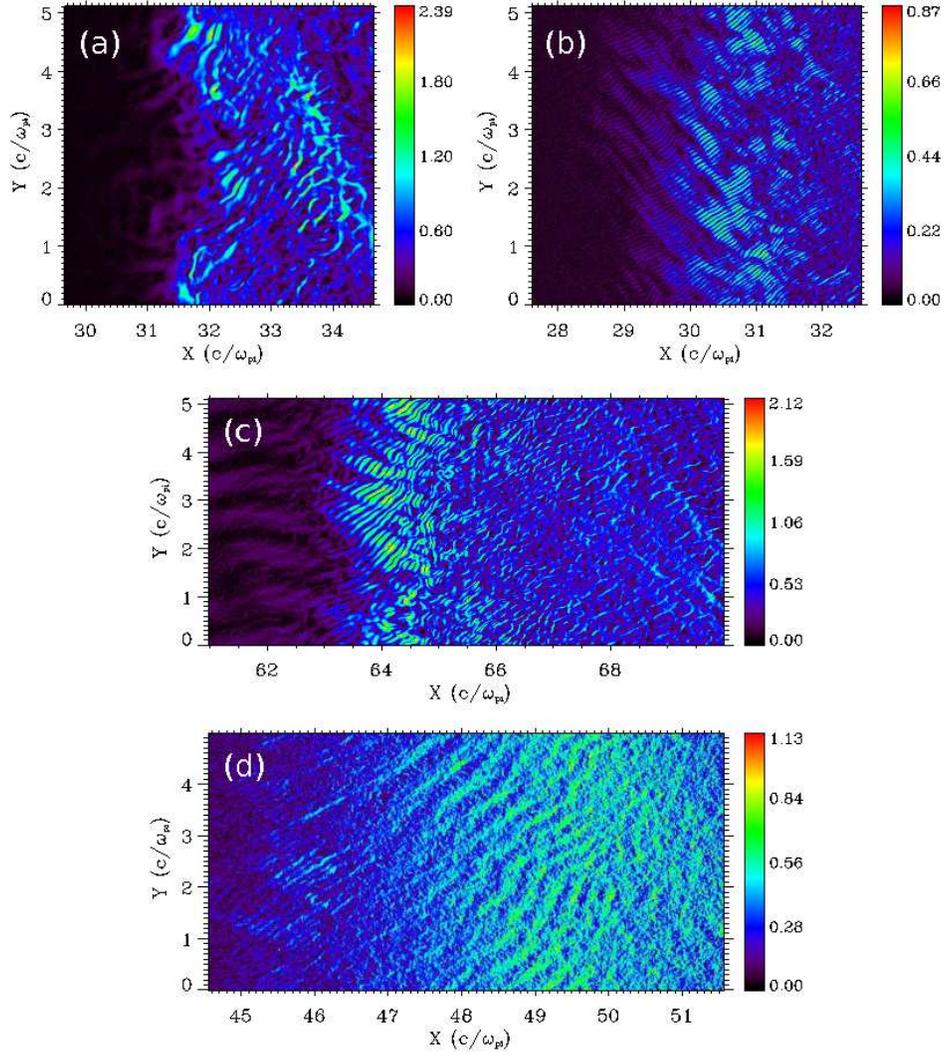}
\caption{Comparison of the electrostatic field strength among simulation runs at $\Omega_{gi}T=6$. Two-dimensional spatial profiles are shown for (a) Run A, (b) Run B, (c) Run C, and (d) Run D.}
\label{fig:est_compar}
\end{figure}

Figure \ref{fig:est_compar} shows a comparison of the electrostatic field strength in the leading edge of the foot. In Run A with $M/m=25$ and $M_A \sim 14$, the strong electrostatic field is excited indicating the Buneman instability (Fig. \ref{fig:est_compar}(a)). The spatial profile is not like a coherent wave form in Run C (Fig. \ref{fig:est_compar}(c)), but consists of isolated peaks distributed in the x-y plane \citep{Amano_Hoshino_2009a}. The amplitude of the electrostatic field reaches $|E_{est}|/B_0 \sim 2.4$ which satisfies the trapping condition of the electron (eq. (\ref{eqn:eb_ratio})). For a reduced mass ratio, the condition of the acceleration toward the relativistic regime (eqs. (\ref{eqn:eb_ratio}) and (\ref{eqn:conversion})) is relaxed to a relatively low Alfv\'en number case, for instance, $M/m=25$ requires $M_A > 10.3$. Run A is indeed the case since $M_A \sim 14$.

In Runs A and C, the Buneman instability is highly activated and the electrons are accelerated in the foot since the trapping condition is satisfied in both simulation runs. However, difference arises in number and the maximum energy of the non-thermal population in the downstream. We have shown in the previous subsection that the second acceleration process is found at the shock surface in Run C which further energizes the pre-accelerated electron from $\gamma\sim3$ up to $\gamma\sim7$ adiabatically. The key for this acceleration is the magnetic and electric fields' profiles around the overshoot. Figure \ref{fig:eb_compar}(a) compares the magnetic field structures in Runs A and C. The peak amplitude of the magnetic field at the overshoot is $B_z/B_0 \sim +8.0$ in Run A which is much smaller than the peak of $B_z/B_0\sim+20.0$ in Run C. The inductive electric field in the ramp found in Run C is not evidently found in Run A (Figures \ref{fig:eb_compar}(b) and \ref{fig:eb_compar}(c)) and thus the betatron acceleration by the reformation process is not effective. The pre-accelerated relativistic electron in Run A cannot gain much energy through the adiabatic accelerations at the shock surface.

If the mass ratio is increased with the Alfv\'en Mach number being fixed, one expects the Buneman instability to be weakly saturated, or even completely stabilized from eqs. (\ref{eqn:unstable_condition}) and (\ref{eqn:conversion}). In Run B, $M/m=100$ and $M_A \sim 16$, and therefore the trapping condition $M_A > 25.8$ is not satisfied while the unstable condition $M_A > 4.2$ is realized (Table \ref{tbl:param}). Figure \ref{fig:est_compar}(b) reflects these conditions that the amplitude of the electrostatic field is $|E_{est}|/B_0 \sim 0.5$ even though the Buneman instability seems to become unstable. Thus the electrons can easily escape from the potential well after moderate energization, and the efficient acceleration to the relativistic regime is not expected in this run.

\subsection{Effect of upstream electron temperature}
Upstream electron temperature also affects the activity of the electrostatic field in the foot; the Buneman instability is expected to be weakly saturated or even completely stabilized in a high electron $\beta_e$ condition. To assess this effect, we simply increased $\beta_e$ from 0.5 in Run C to 4.5 with $M/m$ and $M_A$ being fixed in Run D. In this case, both the trapping condition $M_A > 25.8$ and the unstable condition $M_A > 12.7$ are satisfied (Table \ref{tbl:param}). The resultant electric field profile in the leading edge of the foot is shown in Figure \ref{fig:est_compar}(d). The amplitude of the electrostatic field is so weak ($|E_{est}|/B_0 \sim 0.5$) that any coherent waves are not recognized out of the thermal noise of the hot electron. As in the case of Run B, the non-thermal electrons are hardly found in the downstream.

\subsection{Energy spectra in the downstream region}
\begin{figure}
\plotone{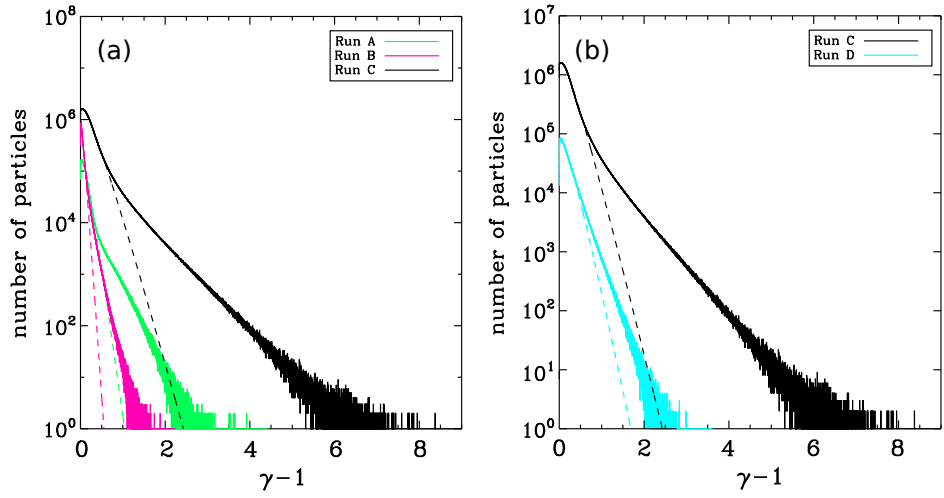}
\caption{Energy spectra of the electron in the downstream region. The results are compared for (a) Run A (green), Run B (magenta), and Run C (black), and (b) Run C (black) and Run D (cyan). Dashed lines are fitted Maxwell distributions.}
\label{fig:edist}
\end{figure}

Lastly, we summarize our simulation runs by showing the energy spectrum of the electron in the downstream region. Particles are sampled in a rectangular region of $0.7\ V_0/\Omega_{gi} \times L_y$ behind the overshoot. Figure \ref{fig:edist}(a) compares the results from Runs A-C. The energy spectrum in Run C exhibits a high energy tail which deviates from the thermal population and reaches $\gamma\sim9$ as was shown in Figure \ref{fig:edist_runc}. A similar non-thermal electron production is observed in Run A in which the Buneman instability is highly activated at the edge of the foot as well. These simulation runs satisfy the trapping condition of eq. (\ref{eqn:eb_ratio}). When the trapping condition does not hold by increasing the mass ratio, the non-thermal population is significantly reduced. This is clearly seen in Run B when compared with Run A. It is also notable that fitted gaussian profiles give the electron temperatures of $T_e/mc^2=0.062$ for Run A and $T_e/mc^2=0.033$ for Run B. Although the Mach number is the same in both runs, the downstream condition is greatly altered by the activity of the Buneman instability in the foot which heats the electrons perpendicular to the magnetic field.

Figure \ref{fig:edist}(b) shows the effect of the upstream electron temperature on the non-thermal electron production by comparing the results from Runs C and D. Even though both the unstable and trapping conditions are satisfied in Runs C and D, the non-thermal population in the downstream in Run D is also significantly reduced. Thus the upstream electron temperature ($\beta_e$) must be another important parameter that determines the efficiency of the electron accelerations. Fitted gaussian profile for Run D gives the electron temperature of $T_e/mc^2=0.11$ which is slightly smaller than in Run C.

The results are summarized in Table \ref{tbl:param} in relation to the upstream conditions.

\section{Summary and Discussion}
We have examined electron acceleration mechanisms at high Mach number shocks by means of two-dimensional Particle-in-Cell (PIC) simulations with various ion-to-electron mass ratios, Alfv\'en Mach numbers, and the upstream electron $\beta_e$. We found electrons are effectively accelerated at a super-high Mach number shock ($M_A\sim30$) with a mass ratio of $M/m=100$ and $\beta_e=0.5$. The electron shock surfing acceleration is found out to be an effective mechanism for accelerating the particles toward the relativistic regime even in two dimensions with the large mass ratio. Buneman instability in the super-high Mach number shock resulted in a coherent electrostatic potential structure which enabled an efficient trapping of the electrons. While multi-dimensionality allows the electrons to escape from the trapping region, they can interact with the strong electrostatic field several times. This multiple interaction enables effective shock surfing accelerations as has been found in the small mass ratio case \citep{Amano_Hoshino_2009a}.

\begin{figure}
\plotone{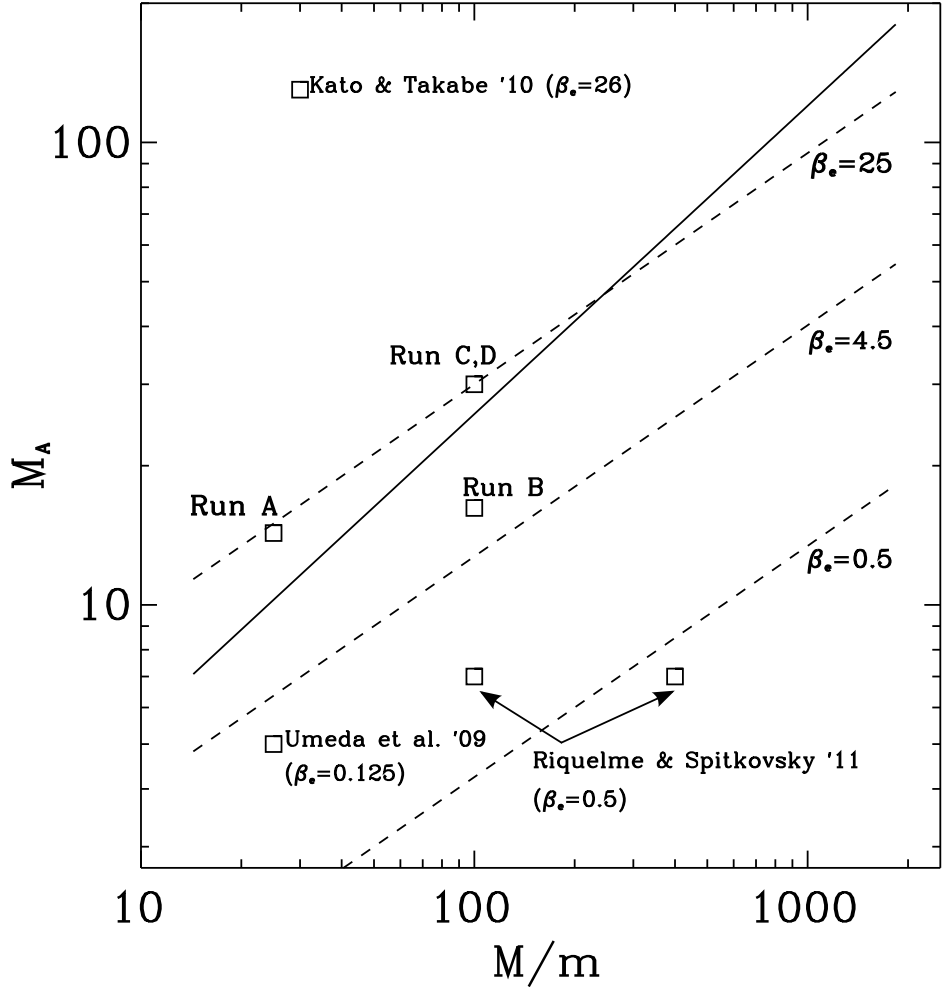}
\caption{Upstream conditions of Runs A-D and the recent two-dimensional PIC simulations as a function of the mass ratio and the Alfv\'en Mach number. Solid line indicates the marginal condition of eq. (\ref{eqn:saturation_level}). Dashed lines indicate the marginal condition of the Buneman instability (eq. (\ref{eqn:unstable_condition})) for selected $\beta_e$ conditions ($\beta_e=0.5,\ 4.5,\ 25$).}
\label{fig:summary}
\end{figure}

The conditions of the electron shock surfing acceleration toward the relativistic regime have been derived from one-dimensional arguments \citep{Papadopoulos_1988,Cargill_Papadopoulos_1988,Ishihara_1980}. Those simple estimations still hold in the present two-dimensional simulations as summarized in Figure \ref{fig:summary}. While all our four simulation runs satisfies the unstable condition of the Buneman instability (dashed lines), the shock surfing acceleration becomes effective in two simulation runs (denoted by ``Run A'' and ``Run C'' in Figure \ref{fig:summary}) whose upstream conditions are also well above the trapping condition of the electron (solid line). A similar aspect holds in recent two-dimensional PIC simulations with different parameters from our simulation runs. For example, the electron shock surfing acceleration does not play important roles in producing non-thermal electrons in shocks with $M_A=7$, $M/m=100,\ 400$ and $\beta_e=0.5$ in which the trapping condition does not meet \citep{Riquelme_Spitkovsky_2011}.

Exception was found for high electron $\beta_e$ cases. In our high $\beta_e$ run (denoted by ``Run D'' in Figure \ref{fig:summary}), the Buneman instability is destabilized in the foot region. However, its peak amplitude is not so large that electrons can be escaped from the trapping region before reaching the relativistic regime. A similar conclusion was also drawn by \citet{Kato_Takabe_2010} in which $M_A\sim130$, $M/m=30$, and $\beta_e=26$ are used (see also Figure \ref{fig:summary}). Although their linear analysis revealed that the foot region in their simulation result was unstable to the Buneman instability, the resultant energy spectrum in the downstream region showed almost a Maxwellian like what we see in our high $\beta_e$ run. These results indicate that we cannot simply understand the high electron $\beta_e$ simulations from Figure \ref{fig:summary} which is based on the linear and nonlinear theories of cold plasma, and detailed analysis of the saturation mechanism of the Buneman instability with finite electron temperature effects is necessary. Extrapolating from the recent and present two-dimensional PIC simulations, Figure \ref{fig:summary} indicates that the electron shock surfing acceleration is an effective mechanism for producing relativistic particles in extremely-high Mach number shocks in supernova remnants (SNRs), provided that the upstream electron temperature is reasonably low.

In addition to the shock surfing acceleration in the foot, the adiabatic accelerations are found to be effective for the pre-accelerated, relativistic electrons in the super-high Mach number shock. The process is a combination of the shock drift acceleration and the betatron acceleration by the time variation of the ramp due to the shock reformation process. Since the mechanism itself is an adiabatic one, the maximum attainable energy  by this process is limited by the amplitude of the magnetic field at the overshoot:
\begin{eqnarray}
\frac{\gamma_{ft}^2}{B_{ft}} &=& \frac{\gamma_{ov}^2}{B_{ov}}, \\
\gamma_{ov} &=& \gamma_{ft}\sqrt{\frac{B_{ov}}{B_{ft}}},
\label{eqn:max_energy}
\end{eqnarray}
where the subscripts ``$ft$'' and ``$ov$'' denotes the foot and the overshoot, respectively. $\gamma_{ft}$ is the energy after the shock surfing acceleration in the foot. The magnetic field strength at the leading edge of the foot does not strongly depend on the Alfv\'en Mach number and our simulation runs exhibit $B_{ft} \sim 2B_0$ which agrees with the multifluid model of \citet{Leroy_1983}. In the Leroy's model, the field strength at the overshoot was also estimated to leading order as
\begin{equation}
B_{ov} \sim \alpha B_0 M_A^{\frac{7}{6}},
\label{eqn:overshoot}
\end{equation}
where we assume $\alpha$ is a constant. This scaling can be applied to our simulation runs up to $M_A=30$. For instance, $B_{ov}=8.0 B_0$ for $M_A=14.4$ in Run A and $B_{ov}=20.0 B_0$ for $M_A=30.0$ in Run C give $\alpha\sim0.4$. Eqs. (\ref{eqn:max_energy}) and (\ref{eqn:overshoot}) are combined to give the maximum energy attained by the adiabatic acceleration around the overshoot as
\begin{equation}
\gamma_{ov} = \sqrt{0.2}M_A^{\frac{7}{12}}\gamma_{ft},
\end{equation}
which suggests that the electrons accelerated in the foot can be further energized by an order of magnitude in extremely-high Mach number shocks in SNRs. Note, however, that the high energy particles reside in the vicinity of the shock surface and their energy is decreased as they are transmitted in the downstream far away from the overshoot owing to the conservation of the magnetic moment. Non-adiabatic processes, such as a pitch angle scattering of the particles by electromagnetic waves at the shock surface, are necessary in order to observe the relativistic electrons in the entire downstream region.

In our two-dimensional purely perpendicular shock simulations with the out-of-plane magnetic field configuration, only drift waves along the shock surface is allowed to exist. Growth of the drift waves (e.g., the lower hybrid drift instability) is observed in the super-high Mach number shock (not shown), however, relationships between the waves and the electron heating or acceleration were not clearly found in the present study. A similar point of view must be introduced if one examines the in-plane magnetic field configuration in which electromagnetic waves are allowed to propagate along the shock surface. In supercritical shocks, two-dimensional hybrid simulations have revealed ion-scale rippled structures at the shock surface associated with strong electromagnetic perturbations, and the ripple is observable only in the in-plane magnetic field configuration \citep{Lowe_Burgess_2003,Burgess_2006}. In this case, the angle between the shock normal and the magnetic field direction changes locally, thereby the reflection rate of the ions is enhanced in some locations feeding more energy to the Buneman instability. Such a connection between the rippled shock structure and the electron acceleration has been reported \citep{Umeda_2009}. The discussions are, however, limited in relatively low Alfv\'en Mach number and small mass ratio cases, and understanding how the ion-scale rippled structure and the electron-scale Buneman instability coexist in high Mach number shocks are remained as a next step.

Note that a two-dimensional simulation can deal with either the two-dimensional electron surfing acceleration in the out-of-plane magnetic field configuration or the rippled structure in the in-plane configuration. In latter case, the surfing acceleration occurs one-dimensionally and the electrons cannot escape from the trapping region. The acceleration efficiency will be different from the present out-of-plane case. Therefore, a unified understanding of electron accelerations in high Mach number shocks will be made possible only by a three-dimensional PIC simulation, which will be reported in near future by virtue of a peta-scale supercomputer system.

\acknowledgments
This work was supported by Grant-in-Aid for Research Activity Start-up 23840047. Numerical computations were partly carried out on Fujitsu FX1 at JEDI of Japan Aerospace Exploration Agency (JAXA), Fujitsu HX600 at Nagoya University, and Cray XT4 at CfCA of National Astronomical Observatory of Japan (NAOJ). 


\clearpage

\clearpage
\begin{table}
\caption{Upstream conditions of simulation runs and resultant electron accelerations \label{tbl:param}}
\begin{tabular}{|c|c|c|c|c|c|c|c|}
\hline
      & $\rm M_A$ & M/m & $\beta_e$ & $\omega_{pe}/\Omega_{ge}$ & eq.(\ref{eqn:unstable_condition}) & eq.(\ref{eqn:saturation_level}) & ele. accel.\\
\hline
Run A & 14.4 & 25  & 0.5 & 10 & 2.1 & 10.3 & BI\\
Run B & 16.2 & 100 & 0.5 & 10 & 4.2 & 25.8 & weak\\
Run C & 30.0 & 100 & 0.5 & 10 & 4.2 & 25.8 & BI + adiabatic\\
Run D & 30.0 & 100 & 4.5 & 10 & 12.7 & 25.8 & weak\\
\hline
\end{tabular}
\label{tbl:param}
\end{table}

\end{document}